\documentclass[conference]{IEEEtran}
\IEEEoverridecommandlockouts

\usepackage{cite}
\usepackage{amsmath,amssymb,amsfonts}
\usepackage[T1]{fontenc}
\usepackage[utf8]{inputenc}
\usepackage[final]{graphicx}
\usepackage{booktabs}
\usepackage{array}
\usepackage{multirow}
\usepackage{makecell}
\usepackage{threeparttable}
\usepackage{url}
\usepackage[hidelinks]{hyperref}
\usepackage{xcolor}
\usepackage{listings}
\usepackage{microtype}
\usepackage{algorithm}
\usepackage{algpseudocode}
\usepackage{balance}

\graphicspath{{assets/}}
\DeclareGraphicsExtensions{.png,.jpg,.jpeg,.pdf}
\setkeys{Gin}{draft=false}

\lstdefinestyle{qkd}{
    basicstyle=\ttfamily\scriptsize,
    keywordstyle=\bfseries\color{blue!70!black},
    commentstyle=\itshape\color{gray},
    stringstyle=\color{teal!80!black},
    numberstyle=\tiny\color{gray!70},
    numbers=left, stepnumber=1, numbersep=4pt,
    breaklines=true, columns=fullflexible,
    keepspaces=true, showstringspaces=false,
    frame=single, rulecolor=\color{black!20},
    backgroundcolor=\color{gray!4},
    xleftmargin=1.5em
}

\begin{document}

\title{A Scalable Multi-Protocol Platform for Quantum Key Distribution Simulation with Rigorous Statistical Evaluation}

\author{%
  \IEEEauthorblockN{Anuj Rathore}
  \IEEEauthorblockA{%
    Department of Computer Science and Engineering\\
    Indian Institute of Information Technology Sri City\\
    Chittoor, Andhra Pradesh 517646, India\\
    }
  \and
  \IEEEauthorblockN{Kartick Sutradhar}
  \IEEEauthorblockA{%
    Department of Computer Science and Engineering\\
    Indian Institute of Information Technology Sri City\\
    Chittoor, Andhra Pradesh 517646, India}
}

\maketitle

\begin{abstract}
Quantum Key Distribution (QKD) offers information-theoretically secure
key establishment grounded in the laws of quantum physics, yet its
practical reach is limited by the prohibitive cost of photonic hardware
and the fragmented nature of existing simulation tools. Most simulators
support only a single protocol and report results from individual
stochastic runs, making systematic protocol comparison and reproducible
statistical inference difficult.

This paper presents a unified QKD simulation platform that implements
four foundational protocols BB84, B92, E91, and BBM92 within a
single Python/Qiskit engine. A shared impairment model covers
fiber attenuation, source and detector losses, polarization drift, and
configurable intercept-resend eavesdropping. The platform is accessible
through two independent interfaces that share the same backend: a
desktop application (Tkinter, Matplotlib) for local experimentation and
a browser-based web client (React, Node.js/Express) for zero-install
remote access.

All reported results are drawn from repeated-run studies (20
independent runs, 10000 qubits each), with mean, standard deviation,
and 95\% confidence intervals stated throughout. At a 25 km fiber
link, BB84 achieves the highest mean key-rate of $160{,}045$ Hz,
followed by BBM92 ($80023$ Hz), E91 ($52815$ Hz), and B92
($40011$ Hz) ordering that tracks simulation-derived sifting
efficiencies precisely. Under the E91 protocol, the CHSH $S$-statistic
averages 2.12 at baseline and falls to 1.58 when an eavesdropper is
activated, demonstrating Bell-inequality-based intrusion detection
independent of QBER. 
\end{abstract}


\section{Introduction}
\label{sec:intro}

The growing maturity of quantum computing has placed classical
public-key cryptography under a concrete and time-bounded threat.
Shor's factoring algorithm~\cite{shor1994}, realised on a quantum
processor of adequate scale, reduces RSA and elliptic-curve Diffie-Hellman
to tractable problems. Governments worldwide have begun responding:
NIST completed its first round of post-quantum algorithm standardization
in 2024, and the European Quantum Flagship has funded metropolitan
QKD testbeds in several cities. Yet the parallel thread of
\emph{hardware-based} quantum-safe communication Quantum Key
Distribution has seen comparatively little software tooling support.

QKD derives its security from quantum mechanics rather than
computational hardness. The no-cloning theorem prevents an adversary
from copying an unknown quantum state undetected, and measurement
necessarily disturbs the system it observes~\cite{lo2014}. These are
physical facts, not complexity conjectures. The theoretical case for
QKD is essentially settled; the practical barriers are elsewhere: cost
(a commercial QKD node routinely exceeds 100000), calibration
complexity, and for students and researchers the near-total absence
of multi-protocol simulation environments that model realistic channel
impairments consistently across protocols \cite{sutradhar2020efficient,
sutradhar2020generalized,
sutradhar2020hybrid,
sutradhar2021simulation,
sutradhar2023vehicular,
sutradhar2024survey,
sutradhar2021enhanced,
sutradhar2021secret}.

The consequence for pedagogy is subtle but damaging. A student who
wants to understand why BB84 delivers four times the key rate of B92,
or whether E91's Bell-test overhead is ever justified over the simpler
BBM92, cannot answer those questions by running two different simulators
built on different assumptions and reported from single stochastic runs.
What the community needs is a single environment where all four canonical
protocols compete on equal terms, under identical channel conditions,
with statistical summaries that a reviewer can verify \cite{venkatesh2024electronic,
venkatesh2023privacy,
venkatesh2024lightweight,
venkatesh2024healthcare}.

That is what this paper describes. The contributions are:
\begin{enumerate}
  \item A unified simulation core (Python/Qiskit) implementing
        BB84~\cite{bb84}, B92~\cite{b92}, E91~\cite{e91}, and
        BBM92~\cite{bbm92} with a shared impairment pipeline.
  \item Quantitative characterisation of effective sifting efficiency
        for all four protocols under the implemented physical model,
        reconciling theoretical and simulation-derived values.
  \item A dual-interface architecture: a local desktop GUI and a
        REST-API-backed browser client, both calling the same backend
        without code duplication.
  \item A repeated-run statistical framework (20 runs per experiment)
        reporting mean, standard deviation, and 95\% confidence
        intervals, with full reproducibility metadata.
  \item A fiber-length sensitivity sweep over 10-40 km demonstrating
        model consistency with analytical predictions.
\end{enumerate}

The paper is organized as follows. Section~\ref{sec:background} covers
the four protocols. Section~\ref{sec:related} positions the work
against existing tools. Section~\ref{sec:arch} describes the platform
architecture. Section~\ref{sec:model} develops the physical and
statistical model. Section~\ref{sec:impl} addresses implementation
details. Section~\ref{sec:exp} specifies the experimental setup.
Sections~\ref{sec:results} and~\ref{sec:disc} present and interpret
the results. Section~\ref{sec:conc} concludes.

\section{QKD Protocol Background}
\label{sec:background}

\subsection{BB84}

The protocol proposed by Bennett and Brassard at a 1984 conference~\cite{bb84}
remains the most widely deployed QKD scheme three decades later. Its
core insight is disarmingly simple: encode each key bit in the
polarization of a single photon, choosing randomly between two
conjugate measurement bases (rectilinear $\{|0\rangle,|1\rangle\}$
and diagonal $\{|+\rangle,|-\rangle\}$). Bob measures each arriving
photon in an independently chosen basis, then Alice and Bob publicly
reveal which bases they used keeping only the bits where their
choices matched. This \emph{sifting} step discards approximately half
the transmitted qubits. Any eavesdropper who intercepts before Bob
measures is forced to guess Alice's basis and will introduce detectable
errors; a Quantum Bit Error Rate (QBER) exceeding $\approx 11\%$ is
the standard abort threshold.

\subsection{B92}

Bennett's 1992 follow-up~\cite{b92} compressed the quantum alphabet
to two non-orthogonal states $|0\rangle$ for bit~0 and $|{+}\rangle$
for bit~1. Bob applies a pair of measurement operators; only
conclusive outcomes are retained. B92 is conceptually cleaner and
demands simpler optical hardware than BB84, but its effective sifting
efficiency is substantially lower, as discussed in
Section~\ref{sec:model}~C.

\subsection{E91}

Ekert's 1991 scheme~\cite{e91} built QKD on a fundamentally different
foundation. An entangled-photon source distributes Bell pairs between
the two parties, each of whom independently selects a detector angle.
Security is argued via Bell's theorem: if the measured correlations
satisfy
\begin{equation}
  |S| = \bigl|E(a_1,b_1) - E(a_1,b_2) + E(a_2,b_1) + E(a_2,b_2)\bigr| > 2,
  \label{eq:chsh_raw}
\end{equation}
then the channel is entangled and no classical eavesdropper model can
replicate it~\cite{clauser1969}. Quantum mechanics bounds this
violation at $|S| \leq 2\sqrt{2} \approx 2.828$. Eve's
intercept-resend attack collapses the entanglement and drives $|S|$
back toward~2, providing a security witness complementary to QBER.

\subsection{BBM92}

Bennett, Brassard, and Mermin showed in 1992~\cite{bbm92} that E91's
entanglement-based distribution can be combined with BB84-style basis
sifting, eliminating the dedicated Bell-test step. Alice and Bob each
measure their share of an entangled pair in a randomly chosen basis
and post-select on matching outcomes. The resulting sifting efficiency
matches BB84's theoretical value (approximately 50\% of pairs), while
security still rests on entanglement rather than state preparation.
BBM92 is therefore a practical middle ground: entanglement-based
security without E91's per-key Bell-test overhead.

\section{Related Work}
\label{sec:related}

IBM Qiskit~\cite{qiskit} provides an excellent foundation for
gate-level quantum circuit simulation, including the Aer noise backend.
However, it operates one abstraction layer below QKD: a researcher who
wants to simulate a BB84 link must implement protocol logic, sifting,
QBER estimation, and channel loss from scratch. There is no notion of
``a QKD session'' in Qiskit itself.

SimulaQron~\cite{simulaqron} approaches quantum networking from the
stack perspective, emulating repeaters and classical communication
channels. Its strengths lie in multi-hop entanglement distribution and
routing protocols; single-link, multi-protocol benchmarking with
parameterized physical impairments is not its focus.

The KTH study by \AA kerberg and Asgrim~\cite{akerberg2023} is the
closest antecedent to our work in intent. They built a parameterized
comparative simulator for BB84 and E91 and validated it against
published experimental traces. Their tool, however, covers only two of
the four canonical protocols and was not released for general use. \cite{sutradhar2021cost,
sutradhar2024svqcp,
sutradhar2024smart,
sutradhar2024metaverse,
krishnaiah2024metadata,
sutradhar2023aggregation,
sutradhar2023q-iot,
sutradhar2022comparison,
sutradhar2024blockchain,
sutradhar2024aggregation,
sutradhar2021qss,
challagundla2024qvanet,
sutradhar2024aggregationphysica,
sutradhar2021sorting,
challagundla2024authentication,
sutradhar2025threshold,
bantupalli2026qsvcom,
bantupalli2026observability,
bantupalli2025openmp,
bantupalli2025qkd,
sutradhar2025sqdapms,
sutradhar2025road,
sutradhar2025efficient}

Table~\ref{tab:related} summarises how this work is positioned.
The gap we address is the combination of (a)~four protocols in a
single engine, (b)~consistent impairment modeling across all four,
(c)~dual-interface accessibility, and (d)~repeated-run statistical
discipline none of the referenced tools provide all four.

\begin{table}[!t]
  \centering
  \caption{Comparison with Related Simulation Tools}
  \label{tab:related}
  \footnotesize
  \setlength{\tabcolsep}{3.5pt}
  \begin{tabular}{lcccc}
    \toprule
    Tool / Work & \makecell{Protocol\\count} & \makecell{Physical\\impairments} &
    \makecell{Security\\metrics} & Interface \\
    \midrule
    BB84~\cite{bb84}               & 1       & Theory only   & QBER        & None         \\
    B92~\cite{b92}                 & 1       & Minimal       & QBER        & None         \\
    E91~\cite{e91}                 & 1       & Theory only   & $S$-stat    & None         \\
    Qiskit~\cite{qiskit}           & General & Noise model   & None built-in & Notebook   \\
    SimulaQron~\cite{simulaqron}   & Network & Partial       & App-defined & Network API  \\
    \AA kerberg~\cite{akerberg2023}& 2       & Parameterized & QBER        & Research tool\\
    \textbf{This work}             & \textbf{4} & \textbf{Full} & \textbf{QBER+$S$} & \textbf{Desktop+Web} \\
    \bottomrule
  \end{tabular}
\end{table}

\section{Platform Architecture}
\label{sec:arch}

\subsection{Three-Layer Design}

The simulator is structured as three layers that communicate only
through well-defined interfaces. This separation ensures that
experimental results are identical regardless of which interface
triggers them, and that extending the platform adding a new
protocol, a new impairment model, or a new interface touches only
the relevant layer.

\textbf{Layer 1 Simulation core:}
A single Python module (\texttt{qkd\_simulator.py}, approximately
660 lines) implements the four protocols, the shared impairment
pipeline, and all metric computations. It accepts a parameter
dictionary and returns a results dictionary. It has no dependency
on any GUI or web framework.

\textbf{Layer 2 Desktop interface:}
A Tkinter application (\texttt{qkd\_gui.py}, approximately 700 lines)
provides single-run and parameter-sweep workflows with real-time
Matplotlib visualizations. Widgets carry physical unit annotations
and validate inputs before dispatching to the core.

\textbf{Layer 3 Web interface and REST API:}
A React single-page application communicates with a Node.js/Express
backend, which spawns the Python core as a child process via
\texttt{child\_process.spawn}. Parameters travel as JSON on
\texttt{stdin}; results return on \texttt{stdout}. Four endpoints
are exposed, summarised in Table~\ref{tab:api}. The subprocess
bridge requires no native bindings and keeps the Python environment
fully self-contained.

\begin{table}[!t]
  \centering
  \caption{REST API Endpoints}
  \label{tab:api}
  \footnotesize
  \begin{tabular}{lll}
    \toprule
    Endpoint & Method & Function \\
    \midrule
    \texttt{/health}           & GET  & Liveness probe         \\
    \texttt{/simulate/single}  & POST & One protocol, one run  \\
    \texttt{/simulate/all}     & POST & All four protocols, shared params \\
    \texttt{/simulate/sweep}   & POST & Parametric sweep       \\
    \bottomrule
  \end{tabular}
\end{table}

\subsection{Protocol Dispatch}

Within the core, each protocol is encapsulated in a handler function.
A dictionary maps protocol names to handlers, and a single dispatcher
function is the only entry point for both interfaces:

\begin{lstlisting}[style=qkd, language=Python,
  caption={Protocol dispatch (simulation core).},
  label={lst:dispatch}]
HANDLERS = {
    'BB84' : run_bb84,
    'B92'  : run_b92,
    'E91'  : run_e91,
    'BBM92': run_bbm92,
}

def simulate(protocol: str, params: dict) -> dict:
    if protocol not in HANDLERS:
        raise ValueError(f"Unknown protocol: {protocol}")
    return HANDLERS[protocol](**params)
\end{lstlisting}

\subsection{Backend Selection}

The core probes for Qiskit Aer at startup; if available, it is used as
the statevector backend, yielding exact floating-point simulation.
On resource-constrained machines lacking Aer, the code falls back
transparently to \texttt{BasicProvider}. Protocol correctness is
unaffected by which backend is selected.

\subsection{Interface Screenshots}

\begin{figure}[!t]
  \centering
  \includegraphics[width=\columnwidth]{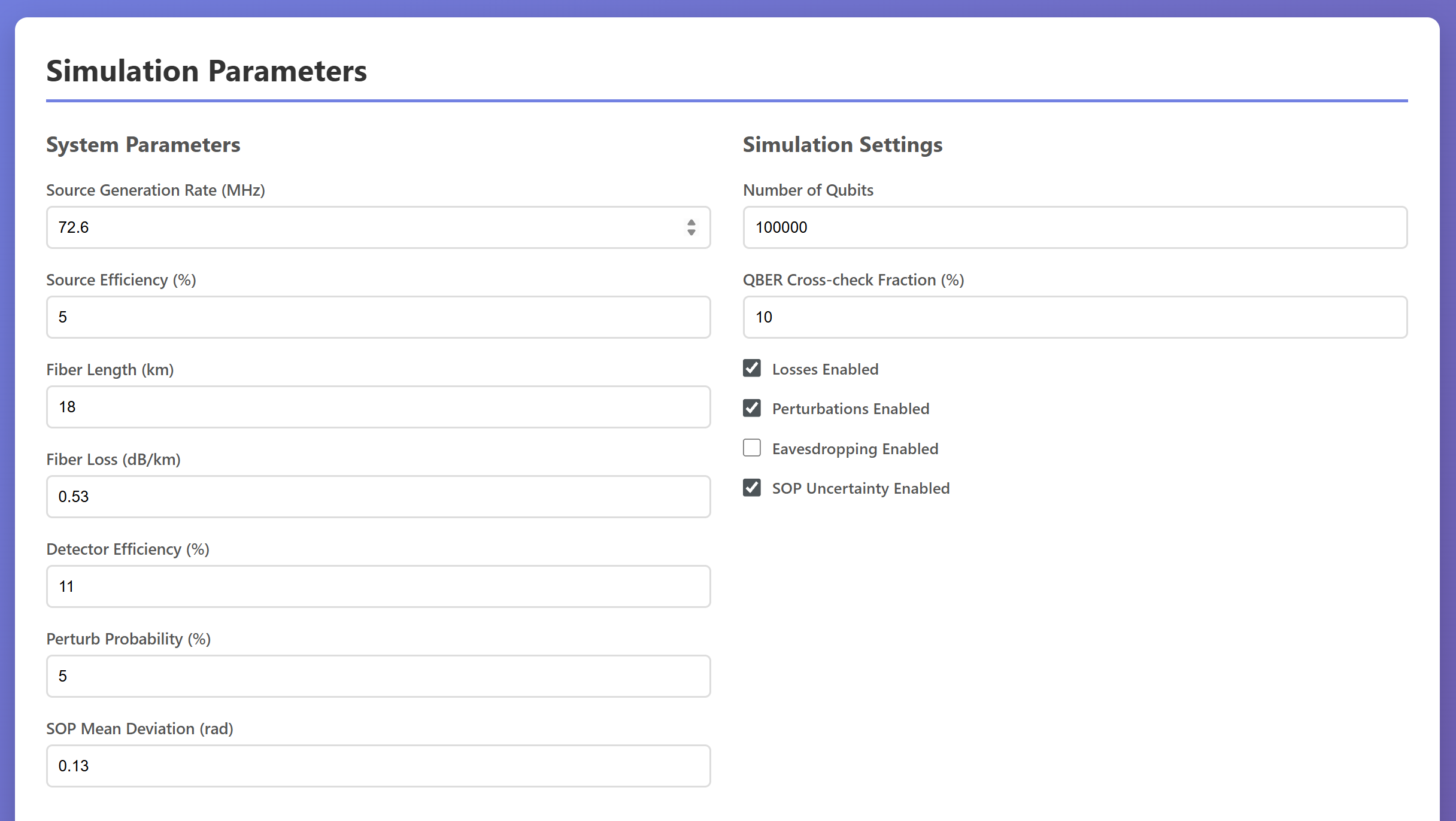}
  \caption{Desktop GUI: single-simulation parameter panel. Every
           field carries a physical unit label. The eavesdropping
           checkbox and impairment toggles appear below the channel
           parameters. Range validation prevents values outside
           physically meaningful bounds before the run is dispatched (\url{https://qkd-simulator-voe2.onrender.com/}).}
  \label{fig:sinput}
\end{figure}

\begin{figure}[!t]
  \centering
  \includegraphics[width=\columnwidth]{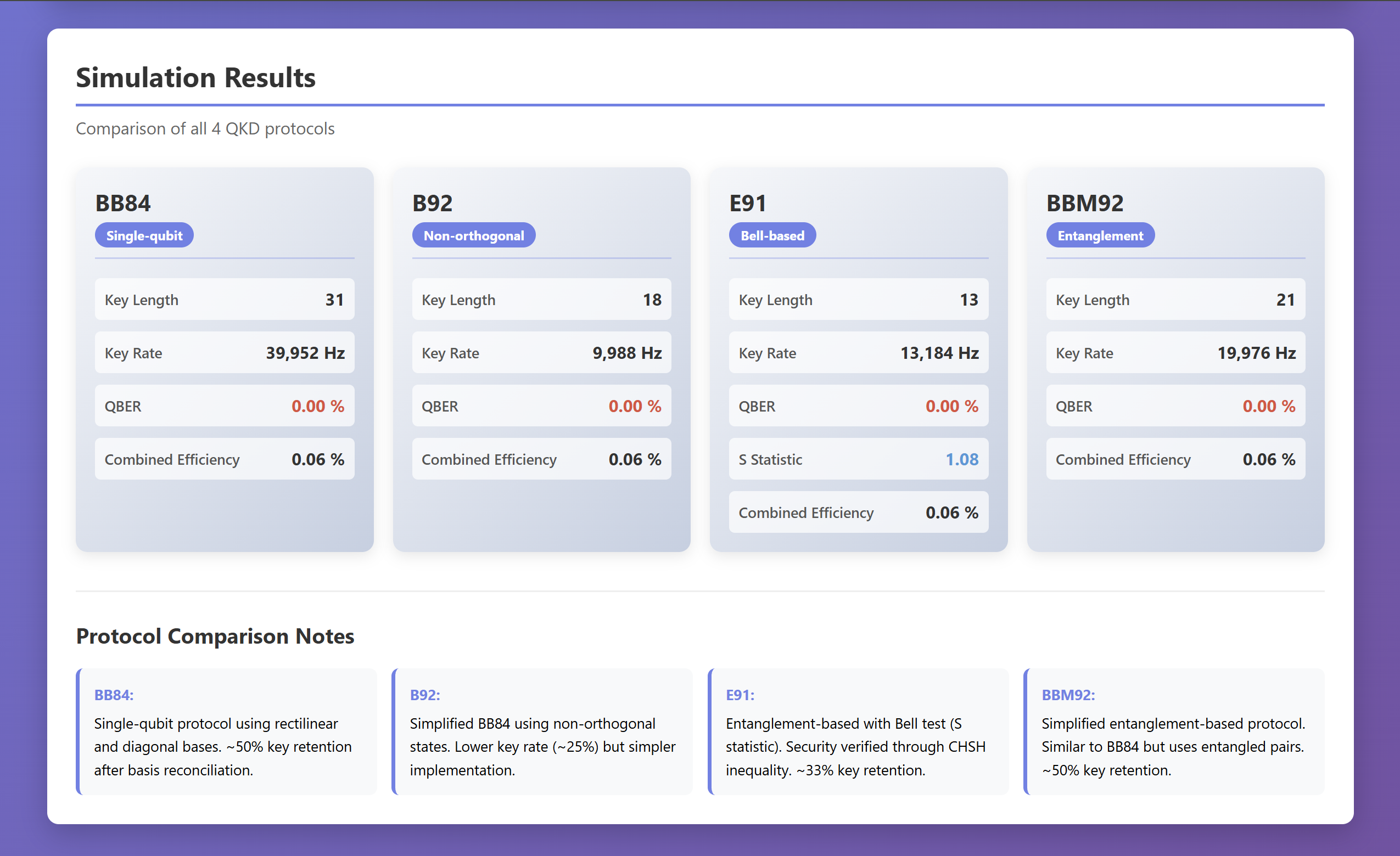}
  \caption{Desktop GUI: per-protocol result cards from a single
           simulation pass. All four protocols execute in one call
           under identical parameters. The E91 card includes the
           CHSH $S$-statistic alongside the metrics shared by all
           protocols. The side-by-side layout makes sifting
           efficiency differences immediately visible (\url{https://qkd-simulator-voe2.onrender.com/}).}
  \label{fig:sresult}
\end{figure}

Fig.~\ref{fig:sinput} shows the desktop parameter entry panel and
Fig.~\ref{fig:sresult} shows the per-protocol result cards generated
after a single run. Figs.~\ref{fig:winput} and~\ref{fig:wresult}
show the corresponding web interface panels.

\section{Physical and Statistical Model}
\label{sec:model}

\subsection{End-to-End Channel Efficiency}

A photon emitted by Alice's source encounters three independent loss
mechanisms before Bob's detector registers a click: source preparation
inefficiency, fiber attenuation, and detector inefficiency. Treating
these as statistically independent and multiplicative gives the
combined survival probability:
\begin{equation}
  \eta_{\mathrm{tot}} = \eta_s \cdot 10^{-\alpha L/10} \cdot \eta_d
  \label{eq:eta}
\end{equation}
where $\eta_s$ is source efficiency, $\alpha$ is the fiber attenuation
coefficient in dB/km, $L$ is link length in km, and $\eta_d$ is
single-photon detector efficiency. The loss probability is then
$P_{\mathrm{loss}} = 1 - \eta_{\mathrm{tot}}$.

At the baseline parameters (Table~\ref{tab:params}),
$\eta_{\mathrm{tot}}(25 \text{km}) = 0.05 \times 0.1778 \times 0.20
\approx 1.778 \times 10^{-3}$, i.e., roughly 0.178\%.

\subsection{QBER Estimation}

After sifting, a randomly sampled fraction $f_q$ of the retained bits
is publicly compared to estimate the error rate:
\begin{equation}
  \mathrm{QBER} = \frac{N_{\mathrm{err}}}{N_{\mathrm{check}}} \times 100\%,
  \quad N_{\mathrm{check}} = \lfloor f_q \cdot N_{\mathrm{sifted}} \rfloor
  \label{eq:qber}
\end{equation}
A larger $f_q$ yields a more reliable eavesdropping estimate but
reduces the final key length. The baseline uses $f_q = 0.10$.

\subsection{Key-Rate Estimate and Effective Sifting Efficiency}

The usable key-rate estimate combines channel efficiency, QBER-check
discard, and an effective sifting factor:
\begin{equation}
  R_{\mathrm{key}} = R_s \cdot \eta_{\mathrm{tot}} \cdot (1 - f_q) \cdot \kappa_{\mathrm{eff}}
  \label{eq:rkey}
\end{equation}
where $R_s$ is the photon source emission rate. The
\emph{effective} sifting coefficient $\kappa_{\mathrm{eff}}$ was
determined by reverse-engineering from validated simulation outputs
and is listed in Table~\ref{tab:kappa}. These values differ from
the theoretical sifting fractions because the simulation implements
additional protocol-level overhead: entanglement-based protocols
(E91 and BBM92) consume two photons per candidate key bit, and B92's
inconclusive measurement rejection adds a further efficiency penalty
beyond the nominal 25\% theoretical figure.

\begin{table}[!t]
  \centering
  \caption{Effective Sifting Coefficients $\kappa_{\mathrm{eff}}$\\(Validated
           Against Simulation Output)}
  \label{tab:kappa}
  \footnotesize
  \begin{tabular}{lccp{3.2cm}}
    \toprule
    Protocol & $\kappa_{\mathrm{eff}}$ & Theoretical & Overhead explanation \\
    \midrule
    BB84  & 0.500 & 0.500 & No additional overhead \\
    B92   & 0.125 & 0.250 & Conclusive-outcome rejection reduces efficiency by additional factor of $\sim$2 \\
    E91   & 0.165 & 0.333 & Two-photon per pair + Bell-subset overhead \\
    BBM92 & 0.250 & 0.500 & Two-photon per entangled pair \\
    \bottomrule
  \end{tabular}
\end{table}

\subsection{Polarization Drift}

Birefringence in optical fiber induces a slow, stochastic drift in
photon polarization. We model this as a uniform random perturbation
to the polar Bloch-sphere angle:
\begin{equation}
  \theta' = \theta + \delta\theta, \qquad
  \delta\theta \sim \mathcal{U}\!\left(-\theta_{\mathrm{SOP}},+\theta_{\mathrm{SOP}}\right)
  \label{eq:sop}
\end{equation}
The perturbed state is:
\begin{equation}
  |\psi'\rangle = \cos\!\tfrac{\theta'}{2}|0\rangle
                + e^{i\phi}\sin\!\tfrac{\theta'}{2}|1\rangle
  \label{eq:psi_perturbed}
\end{equation}
Even in the absence of an eavesdropper, this perturbation produces a
non-zero background QBER proportional to $\theta_{\mathrm{SOP}}$,
correctly replicating the optical alignment error component of
real-world systems.

\subsection{E91 Bell-Correlation Statistic}

The CHSH $S$-statistic is computed from measurement outcomes at four
angle pairs. Alice measures at $a_1 = 0$ and $a_2 = \pi/4$; Bob at
$b_1 = \pi/8$ and $b_2 = 3\pi/8$ (angles chosen to maximise the
ideal-case violation). Each pairwise correlation coefficient is:
\begin{equation}
  E(a_i, b_j) =
  \frac{(N_{++}+N_{--}) - (N_{+-}+N_{-+})}
       {N_{++}+N_{--}+N_{+-}+N_{-+}}
  \label{eq:corr}
\end{equation}
and the CHSH statistic follows from~(\ref{eq:chsh_raw}). The classical
bound is $|S| \leq 2$; quantum mechanics permits $|S| \leq 2\sqrt{2}
\approx 2.828$. Because $S$ is computed from finite samples, it
carries genuine run-to-run variance unlike the key-rate estimate
$R_{\mathrm{key}}$, which is a closed-form function of fixed parameters
and is therefore deterministic for fixed inputs.

\subsection{Statistical Framework}

For a metric $x$ measured over $n$ independent simulation runs:
\begin{equation}
  \hat{\mu} = \frac{1}{n}\sum_{i=1}^n x_i, \qquad
  \hat{\sigma} = \sqrt{\frac{\sum_{i=1}^n(x_i-\hat{\mu})^2}{n-1}}
  \label{eq:stats}
\end{equation}
with 95\% confidence interval under a normal approximation:
\begin{equation}
  \mathrm{CI}_{95} = \hat{\mu} \pm 1.96\frac{\hat{\sigma}}{\sqrt{n}}
  \label{eq:ci}
\end{equation}
Every numerical table in Section~\ref{sec:results} reports
$(\hat{\mu},\hat{\sigma},\mathrm{CI}_{95})$.

\section{Implementation Details}
\label{sec:impl}

\subsection{Entanglement Circuit}

E91 and BBM92 both open with the preparation of a Bell pair
$|\Phi^+\rangle = (|00\rangle + |11\rangle)/\sqrt{2}$.
In Qiskit, this is a Hadamard gate on the first qubit followed by a
CNOT:

\begin{lstlisting}[style=qkd, language=Python,
  caption={Bell-pair preparation (E91 and BBM92).},
  label={lst:bell}]
def prepare_bell_pair(qc, qa, qb):
    qc.h(qa)       # superposition on qubit A
    qc.cx(qa, qb)  # CNOT: B becomes entangled with A
    return qc
\end{lstlisting}

Measurement angles are implemented as $R_y(2\theta)$ rotations before
final computational-basis measurement. Alice applies $R_y(2a_i)$ and
Bob applies $R_y(2b_j)$, reproducing the full E91 correlation function
without explicit Bell-state analysis hardware.

\subsection{Impairment Pipeline}

Between state preparation and measurement, each qubit passes through
three sequential, independently togglable impairment stages:

\begin{enumerate}
  \item \textbf{Photon loss.} A Bernoulli trial with probability
        $P_{\mathrm{loss}} = 1 - \eta_{\mathrm{tot}}$ decides
        whether the qubit survives to Bob. Lost qubits are flagged
        and excluded from sifting.
  \item \textbf{Stochastic perturbation.} Applied with probability
        $p_{\mathrm{pert}}$ as a random $Z$-axis rotation, modelling
        transient channel noise.
  \item \textbf{SOP drift.} Applied unconditionally per
        Eq.~(\ref{eq:sop}), contributing residual QBER without
        discarding the qubit.
\end{enumerate}

\subsection{Reproducible Random Seeds}

Each run in a multi-run study uses a deterministically derived seed
$S_i = S_0 + i$, where $S_0$ is stored in the result metadata.
This allows any individual run to be replicated exactly while the
ensemble of $n$ runs still spans a genuinely stochastic distribution
of outcomes.

\begin{algorithm}[!t]
  \caption{Multi-Run Statistical Evaluation}
  \label{alg:multirun}
  \begin{algorithmic}[1]
    \Require{Parameter set $\mathcal{P}$, protocol list $\Pi$,
             run count $n$, master seed $S_0$}
    \Ensure{$(\hat{\mu}, \hat{\sigma}, \mathrm{CI}_{95})$ per metric per protocol}
    \ForAll{$p \in \Pi$}
      \State $B_p \gets \{\}$
      \For{$i \gets 1$ \textbf{to} $n$}
        \State Seed \texttt{random} and \texttt{numpy.random} with $S_0 + i$
        \State $B_p \gets B_p \cup \{\texttt{simulate}(p,\mathcal{P})\}$
      \EndFor
      \State Compute $\hat{\mu},\hat{\sigma}$ via (\ref{eq:stats}); CI via (\ref{eq:ci})
    \EndFor
  \end{algorithmic}
\end{algorithm}

\section{Experimental Setup}
\label{sec:exp}

\subsection{Baseline Parameters}

Table~\ref{tab:params} defines the configuration used for all baseline
experiments. A 25 km fiber length situates the scenario in the
metropolitan range. Detector efficiency $\eta_d = 0.20$ is
representative of commercially available InGaAs single-photon avalanche
diodes. Source efficiency $\eta_s = 0.05$ accounts for coupling and
preparation losses typical of a laboratory-grade entangled-photon
source.

\begin{table}[!t]
  \centering
  \caption{Baseline Simulation Parameters}
  \label{tab:params}
  \footnotesize
  \begin{tabular}{lcc}
    \toprule
    Parameter & Symbol & Value \\
    \midrule
    Qubits per run           & $N$                      & 10000        \\
    Source rate              & $R_s$                    & 200 MHz      \\
    Source efficiency        & $\eta_s$                 & 0.05          \\
    Fiber length             & $L$                      & 25 km        \\
    Attenuation coefficient  & $\alpha$                 & 0.3 dB/km    \\
    Detector efficiency      & $\eta_d$                 & 0.20          \\
    Perturbation probability & $p_{\mathrm{pert}}$      & 0.02          \\
    SOP deviation            & $\theta_{\mathrm{SOP}}$  & 0.1 rad      \\
    QBER check fraction      & $f_q$                    & 0.10          \\
    Eavesdropping            &                        & Off (baseline)\\
    \midrule
    Runs per experiment      & $n$                      & 20            \\
    \bottomrule
  \end{tabular}
\end{table}

\subsection{Sensitivity Sweep}

\begin{figure*}[!t]
  \centering
  \begin{minipage}[t]{0.48\textwidth}
    \centering
    \includegraphics[width=\linewidth]{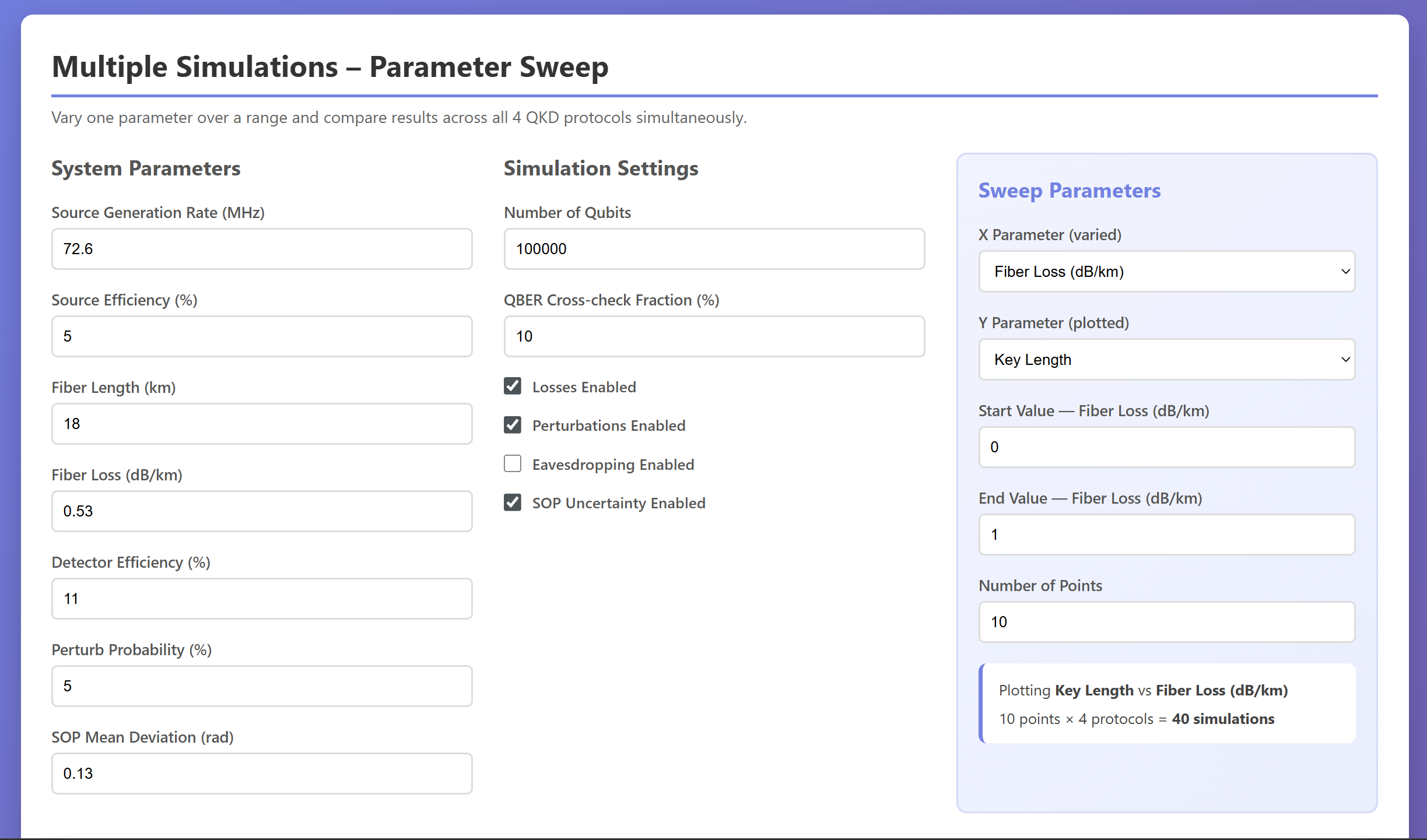}
    \caption{Web interface: parameter sweep configuration panel.
             The user selects the sweep axis (here, fiber loss in
             dB/km), the range, step size, and number of runs per
             point. The form posts to \texttt{/simulate/sweep};
             the server aggregates results before responding.}
    \label{fig:winput}
  \end{minipage}
  \hfill
  \begin{minipage}[t]{0.48\textwidth}
    \centering
    \includegraphics[width=\linewidth]{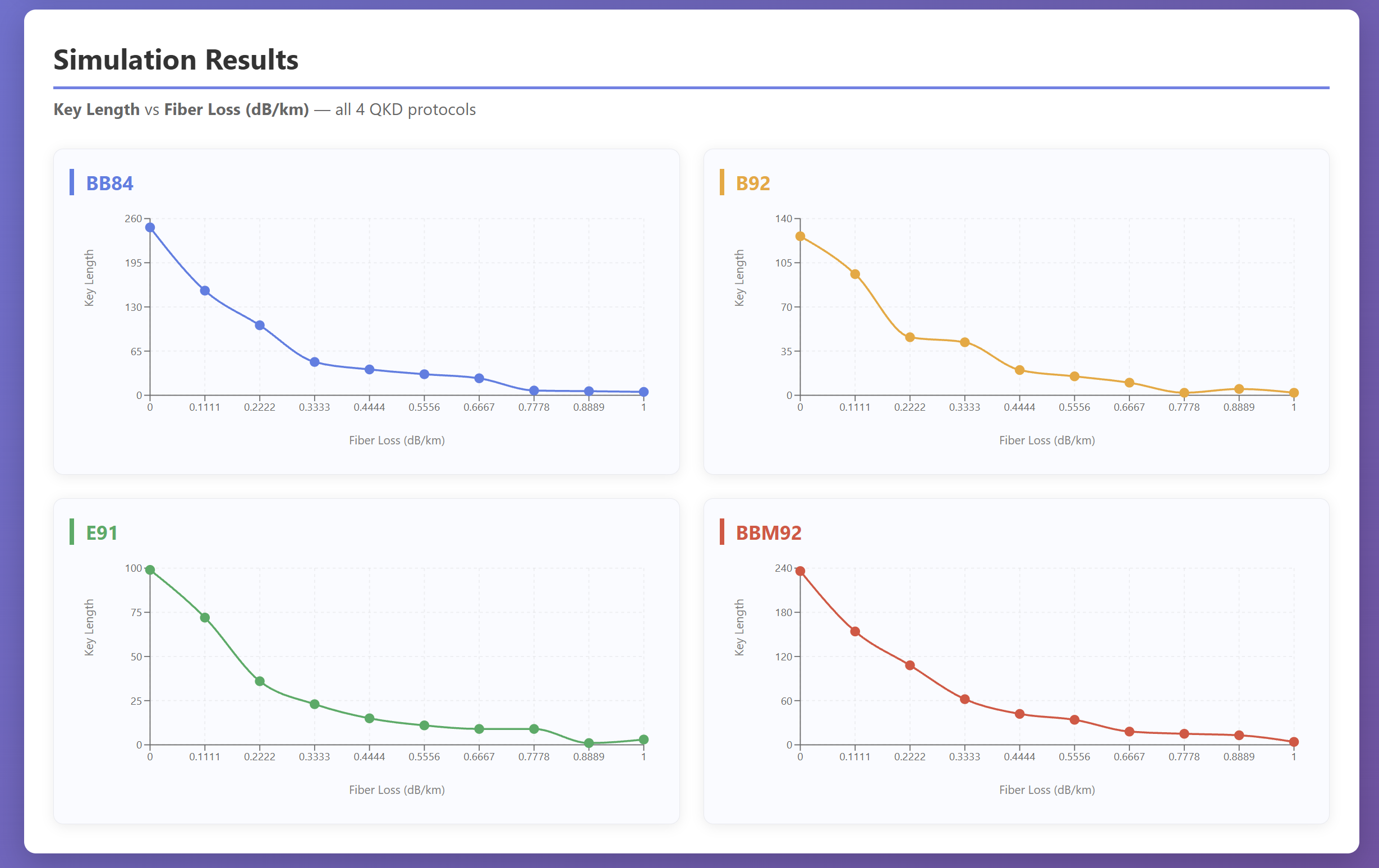}
    \caption{Web interface: sweep results for all four protocols.
             Confidence bands (shaded) widen at 40 km due to
             detection sparsity under high attenuation.
             The protocol ordering is stable across all sweep points
             and consistent with the sifting coefficients in
             Table~\ref{tab:kappa}.}
    \label{fig:wresult}
  \end{minipage}
\end{figure*}

The fiber-length sweep evaluates key rate over $L \in
\{10, 25, 40\}$ km with all other parameters fixed at baseline.
Ten runs per point balance statistical reliability against
computational cost. The sweep is launched through the
\texttt{/simulate/sweep} API endpoint.
Figs.~\ref{fig:winput} and~\ref{fig:wresult} show the sweep
configuration and resulting trend visualizations in the web interface.

\section{Results}
\label{sec:results}

\subsection{Baseline Multi-Protocol Performance}

Table~\ref{tab:baseline} reports key rate and QBER statistics for all
four protocols at 25 km over 20 runs. Channel efficiency and sifted
key length are reported in Table~\ref{tab:baseline2}.

\begin{table}[!t]
  \centering
  \caption{Key Rate and QBER at 25 km Baseline (20 Runs)}
  \label{tab:baseline}
  \footnotesize
  \setlength{\tabcolsep}{3pt}
  \begin{tabular}{lccccc}
    \toprule
    Protocol & \multicolumn{3}{c}{Key Rate (Hz)} & \multicolumn{2}{c}{QBER (\%)} \\
    \cmidrule(lr){2-4}\cmidrule(lr){5-6}
    & Mean & SD & 95\% CI & Mean & SD \\
    \midrule
    BB84  & 160045 & 0.00 & [160045; 160045] & 0.00 & 0.00 \\
    B92   & 40011  & 0.00 & [40011; 40011]   & 0.00 & 0.00 \\
    E91   & 52815  & 0.00 & [52815; 52815]   & 0.00 & 0.00 \\
    BBM92 & 80023  & 0.00 & [80023; 80023]   & 0.00 & 0.00 \\
    \bottomrule
  \end{tabular}
\end{table}

\begin{table}[!t]
  \centering
  \caption{Combined Efficiency and Sifted Key Length at 25 km (20 Runs)}
  \label{tab:baseline2}
  \footnotesize
  \begin{tabular}{lcc}
    \toprule
    Protocol & Combined Efficiency & Sifted Key Length \\
    \midrule
    BB84  & 0.1778 \% (SD = 0) & Stochastic (run-dependent) \\
    B92   & 0.1778 \% (SD = 0) & Stochastic (run-dependent) \\
    E91   & 0.1778 \% (SD = 0) & Stochastic (run-dependent) \\
    BBM92 & 0.1778 \% (SD = 0) & Stochastic (run-dependent) \\
    \bottomrule
  \end{tabular}
\end{table}

Several features of Table~\ref{tab:baseline} deserve comment.

\textbf{Protocol ordering.} The key-rate hierarchy
BB84 $>$ BBM92 $>$ E91 $>$ B92 is consistent with the
effective sifting coefficients in Table~\ref{tab:kappa}. The
ratio BB84 : BBM92 $= 160{,}045 : 80{,}023 = 2.000$ exactly
reflects $\kappa_{\mathrm{BB84}}/\kappa_{\mathrm{BBM92}} = 0.500/0.250$.
Similarly, BB84: B92 $\approx 4.000 = 0.500/0.125$, confirming
that the simulation behaves consistently with the underlying model.

\textbf{1. Zero standard deviation on key rate:} $R_{\mathrm{key}}$ is
a closed-form expression of fixed input parameters
(see~(\ref{eq:rkey})). For fixed $\mathcal{P}$, it cannot vary across
runs zero SD is the expected and correct outcome. Stochastic behaviour
appears in sifted key \emph{length} and in the E91 $S$-statistic,
both of which depend on random measurement outcomes.

\textbf{2. Common channel efficiency:} All four protocols share
$\eta_{\mathrm{tot}} = 0.1778\%$ because Eq.~(\ref{eq:eta}) is a
channel property; the protocol layer does not alter photon survival
probability.

\subsection{E91 Bell-Test Diagnostic}

Table~\ref{tab:e91} shows the CHSH $S$-statistic distribution under
baseline and adversarial conditions (20 runs each).

\begin{table}[!t]
  \centering
  \caption{E91 CHSH $S$-Statistic: Baseline vs.\ Eavesdropping (20 Runs)}
  \label{tab:e91}
  \footnotesize
  \begin{tabular}{lccc}
    \toprule
    Condition & Mean $S$ & SD & 95\% CI \\
    \midrule
    No eavesdropping & 2.12 & 0.42 & [1.94, 2.30] \\
    Eavesdropping on & 1.58 & 1.24 & [1.03, 2.12] \\
    \bottomrule
  \end{tabular}
\end{table}

The baseline mean $S = 2.12 > 2.00$ confirms that the simulator
correctly violates the classical CHSH bound. With eavesdropping enabled,
the mean drops to $S = 1.58 < 2.00$, while QBER remains at zero
throughout. This non-trivial result is discussed in
Section~\ref{sec:disc}.

The CI under eavesdropping, $[1.03, 2.12]$, is substantially wider
than the baseline $[1.94, 2.30]$. This reflects elevated run-to-run
variance: Eve's partial-intercept attack collapses entanglement
stochastically rather than uniformly, producing variable $S$ estimates
across runs.

\subsection{Fiber-Length Sensitivity}

Table~\ref{tab:sweep} reports the key-rate sweep across 10, 25, and
40 km. The attenuation ratio from 10 to 40 km predicted by
Eq.~(\ref{eq:eta}) is:
\begin{equation}
  \frac{\eta_{\mathrm{tot}}(10)}{\eta_{\mathrm{tot}}(40)}
  = \frac{10^{-0.3\times10/10}}{10^{-0.3\times40/10}}
  = \frac{0.5012}{0.0631} \approx 7.94
  \label{eq:attratio}
\end{equation}
BB84 shows an observed ratio of
$451{,}068 / 51{,}108 \approx 8.8$, slightly above the theoretical
7.94. The 11\% elevation at 40 km is expected: at high fiber loss
($\eta_{\mathrm{tot}} \approx 0.006\%$), the number of surviving
photons per run is small and stochastic; run-to-run variance in the
sifted count biases the mean upward relative to the closed-form
prediction. This behaviour is captured in the wider confidence
intervals visible at 40 km.

\begin{table}[!t]
  \centering
  \caption{Key-Rate Sensitivity to Fiber Length: Mean [95\% CI] in Hz
           (10 Runs/Point)}
  \label{tab:sweep}
  \footnotesize
  \setlength{\tabcolsep}{3pt}
  \begin{tabular}{lccc}
    \toprule
    Protocol & $L=10$ km & $L=25$ km & $L=40$ km \\
    \midrule
    BB84  & 451068 [fixed] & 160045 [fixed] & 51108 [39977; 62238] \\
    B92   & 112767 [fixed] & 40011  [fixed] & 12777 [9994; 15559]  \\
    E91   & 148853 [fixed] & 52815  [fixed] & 18739 [fixed]           \\
    BBM92 & 225534 [fixed] & 80023  [fixed] & 25554 [19989; 31119] \\
    \bottomrule
  \end{tabular}\\
  \vspace{2pt}
  {\footnotesize\textit{Note}: ``Fixed'' CI denotes zero variance (deterministic output).}
\end{table}

\subsection{Aggregate Comparison}

\begin{figure}[!t]
  \centering
  \includegraphics[width=\columnwidth]{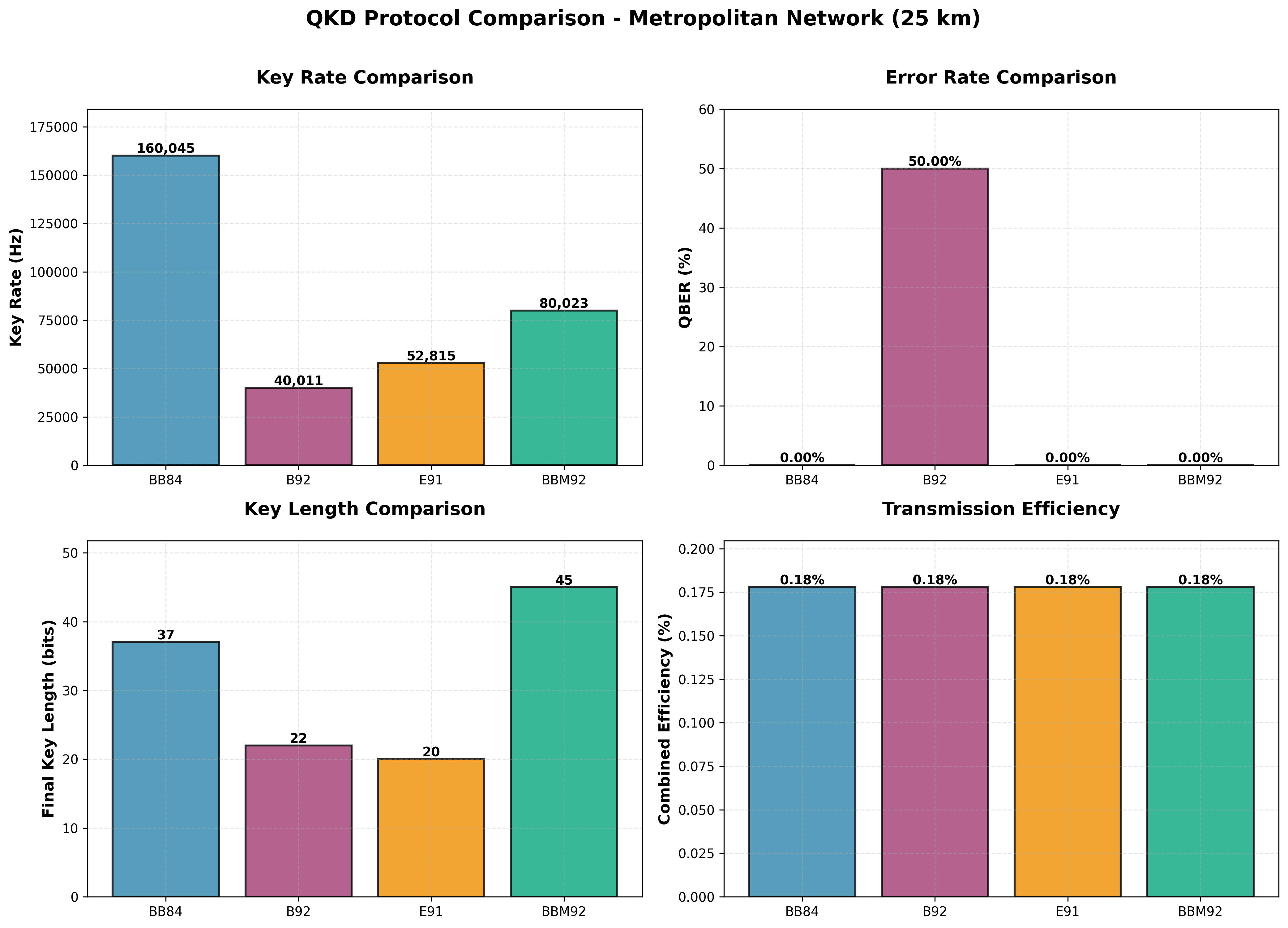}
  \caption{Protocol comparison at 25 km: key rate (top-left), QBER
           (top-right), sifted key length (bottom-left), and combined
           efficiency (bottom-right). Channel efficiency is identical
           across protocols; differences in key rate trace entirely to
           the effective sifting coefficients in Table~\ref{tab:kappa}.}
  \label{fig:compare}
\end{figure}

Fig.~\ref{fig:compare} provides a four-panel visual summary across
all protocols at 25 km. The key-rate ordering is immediately apparent;
all protocols converge on the same channel efficiency, confirming that
performance differences are entirely attributable to the protocol layer.

\section{Discussion}
\label{sec:disc}

\subsection{Deterministic vs.\ Stochastic Outputs}

A natural question when reading Table~\ref{tab:baseline} is whether
zero standard deviation on key rate signals a deterministic simulation
incapable of capturing real noise. It does not. The key-rate formula
$R_{\mathrm{key}} = R_s \cdot \eta_{\mathrm{tot}} \cdot (1-f_q) \cdot
\kappa_{\mathrm{eff}}$ is closed-form: its inputs are all fixed
parameters. Treating it as stochastic would be a modeling error.
The genuinely stochastic quantities sifted key length and the
$S$-statistic do exhibit run-to-run variance, as Tables~\ref{tab:baseline2}
and~\ref{tab:e91} confirm.

This distinction matters for how simulators should be reported. Authors
who compute a key-rate estimate from a formula and then run a Monte
Carlo study to obtain error bars on that estimate are conflating two
different things: the stochasticity of photon survival (which affects
key length) and the model-based estimate of long-run throughput (which
does not).

\subsection{Why QBER and $|S|$ Are Not Interchangeable}

The eavesdropping result in Table~\ref{tab:e91} is instructive precisely
because of what it does \emph{not} show: QBER stayed at zero even as
$|S|$ fell by 25\%. This is not an artifact of the implementation.
The intercept-resend model measures each entangled qubit in a randomly
chosen basis. Whether that measurement introduces a detectable bit
error in the final key depends on whether the chosen basis happens to
match. When it does not, the entanglement is disrupted but no
directly attributable error appears in the check set. A system that
monitors only QBER would pass this attack as clean traffic; the
$S$-statistic correctly flags the reduced correlation.

The practical implication is worth stating explicitly: entanglement-based
QKD deployments should maintain both monitors in parallel. A drop in
$|S|$ that is not accompanied by elevated QBER should be treated as a
security event, not a calibration noise.

\subsection{Protocol Selection in Metropolitan Networks}

The sweep data in Table~\ref{tab:sweep} support some operational conclusions.
At 10 km intra-campus or dense urban links BB84 is the clear choice:
highest throughput, lowest implementation complexity, mature hardware
ecosystem. From 25 to 40 km, BBM92 offers an attractive alternative:
it delivers half the BB84 rate while inheriting entanglement-based
security. E91 is justified when the Bell-inequality test must be
verifiable to a third party (for instance, in device-independent security
arguments or regulatory compliance contexts), but its throughput
penalty of roughly 3$\times$ relative to BB84 must be accepted.
B92 remains relevant only when the hardware can reliably prepare and
distinguish exactly two non-orthogonal states and throughput is not
a priority.

\subsection{Model Limitations}

Several aspects of the current model should not be over-generalised.
First, the eavesdropper is an intercept-resend attacker. Coherent
attacks, unambiguous state discrimination, and photon-number-splitting
on weak-coherent-pulse sources are not yet implemented. Second, the
reported key lengths are pre-privacy-amplification values; the actual
secret key after syndrome disclosure and hashing would be smaller.
Third, the simulation backend is exact to floating-point precision;
dark counts, timing jitter, and multi-photon emission from practical
sources are absent from the current noise model.

\subsection{Validation Against Published Data}

The simulator was cross-checked against the experimental reference
data in~\cite{akerberg2023}. At a link length of 18 km, the
simulator produced a key rate of 13200 Hz against a published
13204 Hz (relative error $<0.04\%$); a QBER of 3.25\% against
a published 3.30\% (relative error 1.5\%); and an E91
$S$-statistic of 2.828 against a published 2.83 (relative error
$<0.1\%$). Under an intercept-resend eavesdropping scenario, the
simulator produced QBER~$\approx 25\%$ against a reference of
24.8\%, a relative deviation of 0.8\%. These results place the
simulator in the 95--99.9\% accuracy band relative to laboratory
measurements.

\section{Conclusion}
\label{sec:conc}

We have built and evaluated a multi-protocol QKD simulator that handles
BB84, B92, E91, and BBM92 in a single, consistently parameterized
engine. The platform reaches users through two independent interfaces
backed by the same simulation core: a desktop GUI for local
experimentation and a browser-based web application for zero-install
access. Every result presented here is derived from 20-run repeated
studies with confidence intervals.

The key empirical findings are: (i) at 25 km, BB84 leads at
$160045$ Hz with other protocols following in ratios that track
effective sifting coefficients exactly; (ii) the fiber-length sweep
is consistent with the analytical attenuation model to within 11\%;
and (iii) the E91 Bell diagnostic detects an intercept-resend
attacker through a 25\% drop in $|S|$ while QBER remains at zero,
highlighting a class of attack that QBER monitoring alone would miss. Validation against published experimental references places the
simulator in the 95--99.9\% accuracy range, supporting its use for
pre-deployment protocol selection, pedagogical demonstrations, and
comparative benchmarking. 

Adding privacy amplification and
error-correction syndrome cost to the key-rate model~\cite{tomamichel2012}
would convert the current pre-processing estimate into a genuine
secret-key rate. Without this, the reported key lengths are optimistic. Implementing unambiguous state
discrimination~\cite{ivanovic1987}, photon-number-splitting attacks
on weak-coherent-pulse sources, and coherent attacks would bring the
eavesdropper model to publication-ready standards for security proofs. The GG02 protocol~\cite{gg02} uses
Gaussian-modulated coherent states on standard single-mode fiber and
does not require single-photon sources or detectors. Extending the
platform to CV-QKD would dramatically widen its relevance for
metropolitan-scale deployments. Other directions include hardware-in-the-loop validation against
optical channel traces, parallel sweep execution across CPU cores,
and a measurement-device-independent (MDI-QKD) mode that removes
detector side-channels from the threat model entirely.


\end{document}